\documentstyle[epsf,seceq,preprint]{jpsj}

\newcommand{\icom}{{\rm i}}
\newcommand{\smint}{\!\int\!}
\newcommand{\smintq}{\smint {\rm d}^2{\mib q}}

\newcommand{\xis}{\xi_\sigma}
\newcommand{\xid}{\xi_{\rm d}}
\newcommand{\zxi}{\xi^{(0)}}
\newcommand{\zxis}{\zxi_\sigma}
\newcommand{\zxid}{\zxi_{\rm d}}
\newcommand{\zsdxi}{\zxi_{\sigma,{\rm d}}}
\newcommand{\sgamma}{\gamma_\sigma}
\newcommand{\dgamma}{\gamma_{\rm d}}
\newcommand{\sdgamma}{\gamma_{\sigma,{\rm d}}}
\newcommand{\zgamma}{\gamma^{\!{}^{(0)}}}

\newcommand{\zsdgamma}{\zgamma_{\sigma,{\rm d}}}

\newcommand{\TPG}{T_{\rm PG}}

\newcommand{\Tc}{T_{\rm c}}

\newcommand{\TKT}{T_{\rm KT}}

\newcommand{\sphi}{{\mib \phi}_\sigma}
\newcommand{\dphi}{\phi_{\rm d}}
\newcommand{\bdphi}{\bar{\phi}_{\rm d}}
\newcommand{\sGamma}{{\mit\Gamma}_\sigma}
\newcommand{\dGamma}{{\mit\Gamma}_{\rm d}}
\newcommand{\schi}{\chi_\sigma}
\newcommand{\dchi}{\chi_{\rm d}}
\newcommand{\uss}{u_{\sigma\sigma}}
\newcommand{\usd}{u_{\sigma {\rm d}}}
\newcommand{\udd}{u_{\rm dd}}

\newcommand{\cs}{c_\sigma}
\newcommand{\cd}{c_{\rm d}}
\newcommand{\TtG}{T_{2{\rm G}}}
\newcommand{\Ec}{E_{\rm c}}

\def\runtitle{Pseudogap and Superconducting Fluctuations in High-Tc Cuprates}
\def\runauthor{Shigeki {\sc Onoda} and Masatoshi {\sc Imada}}

\title{Pseudogap and Superconducting Fluctuations in High-$\Tc$ Cuprates}

\author{Shigeki {\sc Onoda}~\footnote{Email: onoda@issp.u-tokyo.ac.jp} and Masatoshi {\sc Imada}~\footnote{Email: imada@issp.u-tokyo.ac.jp}}
\inst{ Institute for Solid State Physics, 5-1-5 Kashiwa-no-ha, Kashiwa, Chiba, 277-8581}
\recdate{ December 8,1999}

\abst{We analyze pseudogap phenomena widely observed in the underdoped cuprates.  We assume the existence of a strong d-wave pairing force competing with antiferromagnetic(AFM) fluctuations and the formation of flat and damped dispersion around the $(\pi,0)$ and $(0,\pi)$ region as two important elements caused by the proximity from the Mott insulator.  Using the mode-mode coupling theory for the d-wave superconducting(dSC) and AFM fluctuations, we reproduce basic properties of the pseudogap seen in the magnetic resonance, neutron scattering, angle resolved photoemission and tunneling measurements in the cuprates.  Then minimal requirements to understand the pseudogap phenomena are clarified as the above two elements.  A strong competition of the pairing with the antiferromagnetic fluctuations suppresses the transition temperature thereby generates the pseudogap in the underdoped region while the weakness of the AFM fluctuations leads to the absence of the pseudogap at the optimal doping concentration.}

\kword{high-$\Tc$ superconductivity, pseudogap, antiferromagnetism, $d_{x^2-y^2}$-wave superconductivity, Mott transition, strong-coupling superconductor,
mode-mode coupling}

\begin{document}
\sloppy
\maketitle

\section{Introduction} \label{SECTION_intro}

Pseudogap is one of the most remarkable phenomenon in the underdoped region of high-$\Tc$ cuprates. It is observed both in spin and charge
excitations in which gap structure emerges from a temperature $\TPG$ well above the superconducting
transition point $\Tc$.  The gap structure is observed  in various different experimental probes such as NMR
relaxation time, the Knight shift, neutron scattering, tunnenling,
photoemission, specific heat, optical conductivity, and DC
resistivity~\cite{RMP}.  The angle resolved
photoemission spectra
(ARPES)~\cite{LoeserShenDessauMarshallParkFournierKapitulnik1996,photoemission}
have revealed that the pseudogap starts growing first in the
region around $(\pi,0)$ and $(0,\pi)$ from $T=\TPG$ much higher than $T_c$.  In the earlier work~\cite{Gofron},  these momentum regions are known as the region where the quasiparticle dispersion becomes unusually flat and strongly damped.  This gap structure  continuously merges
into the $d_{x^2-y^2}$ gap below $\Tc$.   We call such
$(\pi,0)$ and $(0,\pi)$ momenta ``flat spots" and the region around
them ``flat shoal region".
This region is also known to be particularly important in the understanding of the
metal-insulator transition and its scaling properties~\cite{Imada98,RMPsecIIF11}

One puzzling experimental observation is that the pseudogap
structure appears in $1/T_1T$~\cite{Yasuoka,Y124NMR,Bi2212NMR,Hggap,Hg1223NMR},
while in many cases $1/\TtG$ continuously increases with the decrease
in temperature with no indication of the pseudogap.  In addition, the so called
resonance peak appears in the neutron scattering experiments~\cite{Fong}.  A
resonance peak sharply grows  at a finite frequency below
$\Tc$ with some indications even at $T_c<T<T_{\rm PG}$.  This peak frequency $\omega^\ast$ decreases with lowering
doping concentration implying a direct and continuous 
evolution into the AFM Bragg peak in the undoped compounds.  The neutron and $T_{\rm 2G}$ data support the idea that the AFM fluctuations are suppressed
around $\omega=0$ but transferred to a nonzero
frequency below $\TPG$.  These observations require the framework  
treating the superconducting and AFM fluctuations on an
equal footing.

As we discuss in this paper, the pseudogap phenomena are well
understood as a consequence of two fundamental aspects of the cuprates.  The first is 
proximity effects from the Mott insulator near the metal-insulator 
transition where strong Coulomb repulsion generates a strong and 
critical momentum and energy dependences in electron excitations. 
The coherence temperature (effective Fermi energy) is unusually suppressed due to this proximity. In the momentum space, the flat shoal region appears and this region determines the basic character of the metal-insulator transition.  The strong correlation effects appear most critically in this region with formation of flattened and strongly damped dispersion.  Because of its flatness with diverging density of states, the doping effects are determined predominantly by this region.  The flat shoal region has a fundamental importance also in clarifying the mechanism of the pseudogap since the ARPES result shows that the pseudogap is first formed from this region.  

The second fundamental aspect in the cuprates is the strong coupling nature of the pairing interaction.  The short coherence length observed in the cuprate superconductors implies that the effective Fermi energy $E_F$ is comparable or even smaller than the energy scale of the pairing interaction in contrast with the conventional BCS superconductors.  This is indeed a natural consequence of the suppressed coherence and  $E_F$ for a metal near the Mott insulator when the pairing force is kept constant.  The appearance of the pseudogap region characterized by a separation of superconducting transition temperature $T_c$ and the onset of pairing fluctuation is naturally understood from this strong coupling character.  In this paper, we see that this separation is strongly enhanced by the repulsive mode-mode coupling between dSC and AFM fluctuations because the AFM fluctuations suppress superconducting $T_c$.  

To reach full understanding of the pseudogap phenomena, we need satisfactory  descriptions of both of the above two aspects, although even a complete description of each aspect alone has never been given in the literature.  The formation of the flat shoal region is observed in numerical studies~\cite{Imada98,Assaad98}  while it is not well reproduced in self-consistent treatments by the diagrammatic approaches.  Concerning the second aspect, several strong coupling approaches for the superconducting phase in the literature have not seriously treated neither the competition of the pairing interaction with the antiferromagnetic fluctuations nor the suppression of the coherence temperature.  These so far ignored elements are actually crucial elements in realizing a region of strong pairing fluctuation above the suppressed $T_c$.   Furthermore, under the serious competition of the two fluctuations, the origin of the strong pairing force has not been clarified yet.  This is presumably because the pairing force itself results from the incoherent high-energy excitations under the proximity effect from the Mott insulator while microscopic theory for such incoherent part is not fully developed.  This difficulty indeed becomes clear when we see the results of this paper, where the antiferromagnetic fluctuations must be repulsive with the pairing force at low energies in reproducing the pseudogap formation and a rather high-energy incoherent excitations are required for the origin of the pairing force.

 Keeping the present stage of the above understanding in mind, we develop a theoretical framework of the pseudogap phenomena to account for all of the basic experimental results.    The scope of this paper is not so ambitious that a full microscopic theory is constructed.  In this paper we rather aim at giving a framework where the various experimental results are reproduced from a theory starting from the observation of the above two aspects, namely the strong coupling nature of pairing and critically strong momentum dependence in the quasi-particle excitations.   In this paper, irreducible pairing interaction is rather given as input and assumed to exist. We also pick up enhanced contributions from the flat shoal region to the dSC and AFM susceptibilities and impose a cutoff in the integral over momentum space to mimic the dominant contribution of this region, although the flatness and the damping are not completely expressed in this scheme.   We then construct a mode-mode coupling theory for both the d-wave pairing and antiferromagnetic fluctuations on an equal footing.  We will show that such minimal requirements are enough to reproduce the basic experimental results of the pseudogap phenomena~\cite{Onoda991,Onoda992}.  

Although the existence of the flat shoal region plays main role for the criticality of the metal-insulator transition and the formation of the pseudogap, a subtlety arises in some physical quantities for the role of the other region as $(\pi/2,\pi/2)$ point. In fact, the DC transport and damping of the magnetic excitations could substantially be influenced from doped holes in the other dispersive region.  Although the transition to the Mott insulator is not accompanied by the critical behavior of the relaxation time $\tau$ but by $\tau$-independent quantities as the Drude weight and the compressibility, the noncritical quantities such as the DC transport and the magnetic relaxation may sensitively depend on $\tau$.  This is particularly true for the damping of the magnetic excitations under the pseudogap formation.   If contributions from the $(\pi/2,\pi/2)$ region would be absent, the damping of the magnetic excitation would be strongly reduced when the pseudogap is formed around $(\pi,0)$.  However, under the pseudogap formation, the damping can be determined by the Stoner continuum generated from  the  $(\pi/2,\pi/2)$ region and can remain constant.  This process is in fact important if the quasiparticle damping around the  $(\pi/2,\pi/2)$ region is large as in the case of La 214 compounds.  Since the whole momentum dependence of the quasiparticle damping is not easy to derive in the present stage, and the damping of the magnetic excitations are determined from rather complicated combination from the both flat and dispersive regions, in this paper, we leave the damping of the magnetic excitations as an input from the outside of the framework based on phenomenological grounds.  The formation of the pseudogap itself is a rather universal consequence of the strong coupling superconductors.  However, as we see below, the actual behavior may depend on this damping.  For example, we show below that the damping generated by the  $(\pi/2,\pi/2)$ region sensitively destroy the resonance peak structure observed in the neutron experimental results.  

\section{Mode-mode coupling treatment} \label{SECTION_Seff}

We consider a 2D strongly correlated electron system and treat
AFM and pairing fluctuations simultaneously~\cite{Onoda991}.
Following the argument in \S 1, we represent
the partition function of the system by the functional integral over
both of the AFM and $d$SC auxiliary fields, $\sphi$ and $\dphi$ introduced by the Stratonovich-Hubbard transformation.

After integrating out the fermions degrees of freedom, the following effective action is obtained,

\begin{full}
\begin{eqnarray}
S &=& S^{(0)} + S_\sigma^{(2)} + S_{\rm d}^{(2)}
+ S_{\sigma\sigma}^{(4)} + S_{\rm dd}^{(4)} + S_{\sigma {\rm d}}^{(4)},
\label{S} \\
S_\sigma^{(2)} &=& \beta\sum_n\smintq A_{\sigma}\schi^{-1}(\icom\omega_n,{\mib q})
\sphi(\icom\omega_n,{\mib q})\!\cdot\!\sphi(-\icom\omega_n,-{\mib q}),
\label{S_sigma2} \\
S_d^{(2)} &=& \beta\sum_n\smintq A_d\dchi^{-1}(\icom\omega_n,{\mib q})
\bdphi(\icom\omega_n,{\mib q})\dphi(\icom\omega_n,{\mib q}),
\label{S_d2} \\
S_{\sigma\sigma}^{(4)} &=& \beta\uss\hspace{-10pt}\sum_{n_1,n_2,n_3}\!
\smintq_1\smintq_2\smintq_3
\sphi(\icom\omega_{n_1},{\mib q}_1)\!\cdot\!\sphi(\icom\omega_{n_2},{\mib q}_2)
\sphi(\icom\omega_{n_3},{\mib q}_3)\!\cdot\!\sphi(\icom\omega_{n_4},{\mib q}_4),
\label{S_sigma4} \\
S_{\rm dd}^{(4)} &=& \beta\udd\hspace{-10pt}\sum_{n_1,n_2,n_3}\!
\smintq_1\smintq_2\smintq_3
\bdphi(-\icom\omega_{n_1},-{\mib q}_1)\dphi(\icom\omega_{n_2},{\mib q}_2)
\bdphi(-\icom\omega_{n_3},-{\mib q}_3)\dphi(\icom\omega_{n_4},{\mib q}_4),
\label{S_d4} \\
S_{\sigma {\rm d}}^{(4)} &=& 2\beta\usd\hspace{-10pt}\sum_{n_1,n_2,n_3}\!
\smintq_1\smintq_2\smintq_3
\sphi(\icom\omega_{n_1},{\mib q}_1)\!\cdot\!\sphi(\icom\omega_{n_2},{\mib q}_2)
\bdphi(-\icom\omega_{n_3},-{\mib q}_3)\dphi(\icom\omega_{n_4},{\mib q}_4),
\label{S_sigmad4}
\end{eqnarray}
\end{full}

\noindent
where $\sphi$ is the three-component vector field corresponding to the spin,
and $\bdphi$ and $\dphi$ are the pairing fields creating and annihilating
a pair of electrons, respectively.
$\beta$ is the inverse temperature and $\omega_n$ and ${\mit\Omega}_m$ are
bosonic and fermionic Matsubara frequencies.
Here 
\begin{eqnarray}
\lefteqn{\hspace{-11.5mm}\schi(\icom\omega_n,{\mib q})\!=\!A_{\sigma}\hspace*{-1mm}\left(\zxis{}^{-2}\hspace*{-1mm}+\hspace*{-1mm}({\mib q}-{\mib Q})^2\hspace*{-1mm}+\hspace*{-1mm}\frac{\sgamma|\omega_n|}{\cs^2}\hspace*{-0.5mm}+\hspace*{-0.5mm}\frac{\omega_n^2}{\cs^2}\right)^{-1}\hspace*{-4mm},}
\label{chi_sigma} \\
&&\hspace{-11.5mm}\dchi(\icom\omega_n,{\mib q})\!=\!A_d\left(\zxid{}^{-2}+{\mib q}^2+\frac{\dgamma|\omega_n|}{\cd^2}
+\frac{\omega_n^2}{\cd^2}\right)^{-1},
\label{chi_d}
\end{eqnarray}
are AFM and $d$SC dynamical susceptibilities,
${\mib q}_4=-{\mib q}_1-{\mib q}_2-{\mib q}_3$, $n_4=-n_1-n_2-n_3$,
and ${\mib Q}$ is the AFM ordering wave vector.
The bare correlation lengths $\zxis{}$ and $\zxid{}$ are determined from the inverse of the bare susceptibilities calculated from the bare dispersion $\varepsilon_{\bf k}$.  Here, we
take the Hubbard model with nearest-neighbor transfer $t$ and the next-nearest-neighbor
transfer $t'$, which leads to $\varepsilon_{\bf k} = -2t({\rm cos}k_x
+{\rm cos}k_y)-4t'({\rm cos}k_x{\rm cos}k_y+1)-\mu$.  The chemical potential  $\mu$ is measured from the flat spots. 
We take the form 
\begin{eqnarray}
\lefteqn{\hspace{-15mm}\zxis{}^{-2}\approx
\!1\!-\!\frac{|\sGamma|}{t}\log\frac{\Ec}{max\{\mu,t',T\}}
\log\frac{\Ec}{max\{\mu,T\}}} \label{RPA} \\
&&\hspace{-15mm}\zxid{}^{-2}\approx\!1\!-\!\frac{|\dGamma|}{\sqrt{t^2-4t'^2}}
\log\frac{\Ec}{T}\log\frac{\Ec}{max\{\mu,T\}} \label{TMA}
\end{eqnarray}
which is valid for the contributions from the $(\pi,0)$ and $(0,\pi)$ regions.
Here $\Ec$ is a ultraviolet cutoff in the energy scale of the
bandwidth and we have included coefficients of the double logarithms
into the original AFM and dSC Gaussian coupling coefficients $\sGamma$ and $\dGamma$. In the above forms, the role of the flat shoal region is emphasized and selectively picked up by a cutoff imposed in the momentum space.

 We note that although we calculate the bare correlation length from the Hubbard model, our scheme is beyond the scope of it  because we introduced the AFM and dSC coupling constants $u$, $\Gamma$, $A_{\sigma}$ and $A_{d}$ as phenomenological parameters.

In this paper we restrict ourselves to the AFM ordering vector
at the commensurate value $(\pi,\pi)$.
We also neglect possible long-range features of Coulomb interaction
which may lead to gapful $d$SC excitations instead of the Goldstone
mode even in the $d$SC ordered state, as in the $s$-wave SC state.
We have confirmed that the susceptibility $\dchi$ taken to satisfy the 
Anderson-Higgs mechanism instead of (\ref{chi_d}) dose not alter the
qualitative feature of the pseudogap behavior obtained in this paper.
The phase excitations (Higgs bosons) are not treated separately from
the amplitude modes.
The velocity of spin and pairing collective modes are
denoted by $\cs$ and $\cd$. The damping constants are given by 
$\dgamma$ and $\sgamma$.

Following  the argument in \S 1 for the damping of the AFM fluctuations, we introduce a phenomenological form of $\sgamma$ and $\dgamma$.
The origin of $\gamma$ is mainly from
continuum of the Stoner excitations and the amplitude strongly depends
on low-energy quasiparticle excitations.  This low-energy
part of damping becomes negligible if some kind of long-ranged order
appears. It may also be suppressed if the correlation length gets
longer. When only one type of fluctuations with the correlation length
$\xi$ exists, a plausible dependence for long $\xi$ would be
$\gamma=\zgamma \xi^{-\varphi}$.  In case of the enhanced d-wave correlation length, however, the situation is not so simple, because the low-energy excitations around $(\pi/2,\pi/2)$ are not suppressed due to the nodal structure of the d-wave gap.  To get qualitative results, these suggest us a rough form for the damping as $\gamma \propto \gamma_1/(\xi_{\sigma}^{-\varphi}+\xi_d^{-\varphi})+\gamma_2/\xi_{\sigma}^{-\varphi}$. Here the first term represents the contribution from the $(\pi,0)$ region and the second term is from the $(\pi/2,\pi/2)$ region.  Since the damping does not have critical change in the pseudogap region, the above rough form may be a good starting point to get an idea for its role.  Depending on the relative amplitude of the first and second terms, we may take only the dominant term between these two.  In the pairing dominant region as in the subject of this paper, these two choices may also be  expressed simply as 
\begin{equation}
\sdgamma=2\zsdgamma/(\xis^{\varphi}+\xid^{\varphi}),
\label{damp_z=1whole}
\end{equation}
 where we should take $\varphi=0$ in the latter choice for the $(\pi/2,\pi/2)$ contribution.  
In terms of bosonic excitations,
the relaxation times of collective modes should be determined
by the time necessary to propagate the scale of the longest
correlation length, because the damping is not effective as far as the 
excitations are propagating inside such an ordered domain.
Thus we take $\varphi=1$ for the former choice.  The case $\varphi=1$ represents the one where the damping at $(\pi,0)$ is overwhelming over the $(\pi/2,\pi/2)$ region thus is generally adequate for the underdoped region.  The exception is the La-based 214  compounds, where the quasiparticle damping around $(\pi/2,\pi/2)$ is unusually large presumably because of charge ordering fluctuations.  The optimally doped compounds are rather expressed by $\varphi=0$ because the relatively weak damping around $(\pi,0)$ does not allow us to neglect the contribution from the $(\pi/2,\pi/2)$ region any more.  

Although we take different choices, $\varphi=0$ or 1 for optimal and underdoped regions, respectively, we note that the form for the damping does not alter the formation of the pseudogap itself.  The pseudogap formation is a consequence of large d-wave coupling constant competing with AFM fluctuations under unusually suppressed coherence temperature.  We will see later that the pseudogap appears only in the underdoped situation in our scheme.  This is simply because the mutual competition between dSC and AFM fluctuations are severe there and additionally the coherence temperature becomes low.  The damping form is crucial only for the appearance of the resonant peak in the underdoped region.  If we take $\varphi=0$ instead of our present choice $\varphi=1$ in the underdoped region, the pseudogap survives but the resonant peak would not appear.  

Using the effective action obtained above, we perform
renormalization process for the mode-mode coupling terms.  This is a
similar procedure to the SCR theory developed by Moriya and
coworkers for spin fluctuations~\cite{SCR}.
In our case the mode-mode coupling terms consist of those between AFM
and AFM fluctuations, $d$SC and $d$SC fluctuations, and AFM and $d$SC
fluctuations. Following the mode-mode coupling scheme, 
$\xis$ and $\xid$ are determined selfconsistently from

\begin{full}
\begin{eqnarray}
\xis^{-2}&=&\zxis{}^{-2}+
\!\int_0^{2\Ec}\!\frac{{\rm d}\omega}{\pi}\!\int_0^1\!\frac{{\rm d}^2{\mib k}}{(2\pi)^2}\coth\frac{\omega}{2T}
\left[\uss{\rm Im}\schi(\omega,{\mib k})
+\usd{\rm Im}\dchi(\omega,{\mib k})\right],
\label{SCRs} \\
\xid^{-2}&=&\zxid{}^{-2}+
\!\int_0^{2\Ec}\!\frac{{\rm d}\omega}{\pi}\!\int_0^1\!\frac{{\rm d}^2{\mib k}}{(2\pi)^2}\coth\frac{\omega}{2T}
\left[\usd{\rm Im}\schi(\omega,{\mib k})
+\udd{\rm Im}\dchi(\omega,{\mib k})\right],
\label{SCRd}
\end{eqnarray}
\end{full}
where in the susceptibilities $\chi$, the bare correlation lengths $\xi_{\sigma}^{(0)}$ and $\xi_{d}^{(0)}$ in (2.7) and (2.8) are replaced with the renormalized ones without $(0)$. 

In this formalism for two-dimensional systems,
the system can be ordered only at $T=0$.
It agrees with the Mermin-Wagner theorem~\cite{MerminWagner66} for the
AFM order.
For the $d$SC, however, the K-T transition at nonzero temperatures is not
reproduced since the SCR theory cannot describe such topological K-T transition at
$\TKT$.
However, the renormalized superfluid stiffness determines the temperature
scale where the pairing correlation length starts growing strongly.
Since $\TKT$ is of the order of the stiffness
according to the K-T theory~\cite{K-Ttransition},
$\TKT$ is close to this crossover temperature $T_*$
in our theory, below which the spin correlation length starts
decreasing.  In our analysis, we take this temperature
scale as the signature of the K-T transition.

For our calculation, we choose two sets of parameter values. In one, parameter values for typical underdoped compounds such as YBa$_2$Cu$_3$O$_{6.63}$ and YBa$_2$Cu$_4$O$_{8}$ are taken 
and for the other, typical optimally doped case such as YBa$_2$Cu$_3$O$_{7}$ is considered.
For details of the determination of the parameters readers are referred to the paper~\cite{Onoda991,Onoda992}.

\section{Results} \label{SECTION_results}

\begin{figure}[htb]
\epsfxsize=7.8cm
$$\epsffile{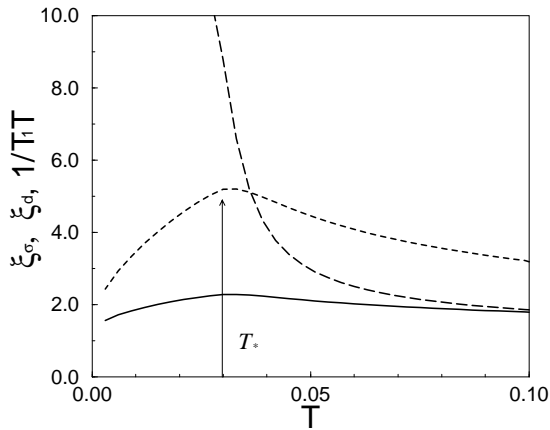}$$
\vspace{-0.7cm}
\caption{The spin correlation length (solid line), $1/T_1T$ (dashed line) normalized by its value at $T=t$, and $d$SC correlation length (long-dashed line) plotted as functions of temperature. The parameter values are for the optimally doped cuprates and given as $|\sGamma|=|\dGamma|=0.15$, $\uss=2.6$, $\usd=1.2$, $\udd=1.8$, $\mu=0.03$ and $t'=-0.03$.  This choice corresponds to YBa${}_2$Cu${}_3$O${}_7$.}
\label{Figxi-phi0}
\end{figure}
First, we discuss results for optimally doped systems. The parameter values
are chosen from an optimally doped compound, YBa${}_2$Cu${}_3$O${}_7$~\cite{Yasuoka} and $\varphi=0$ is also taken.
We plot the calculated $d$SC correlation length $\xid$, $1/T_1T$ and
spin correlation length $\xis$ in Fig.~\ref{Figxi-phi0}. 
In this case, the spin correlation length first increases down to the
temperature $T_*$ and then decreases with further decrease in
temperature. 
It makes a crossover between the regime $T<T_*$ dominated
by the $d$SC renormalized classical fluctuations and the thermally
fluctuating regime $T>T_*$.

The spin-lattice relaxation rate $1/T_1$ of a ${}^{63}$Cu nuclei
is evaluated from the momentum sum of the imaginary part of the dynamical spin susceptibility.
Here we have not considered the ${\bf k}$-dependence of the nuclear
form factor seriously, because it does not alter the basic feature.
We see that $1/T_1T$ has basically the same
temperature dependence as $\xis^2$.
Our present result for
$\varphi=0$ is also similar to the results for Tl${}_2$Ba${}_2$CuO${}_6$,
or HgBa${}_2$CuO${}_{4+\delta}$~\cite{Hg1201NMR}.
Our results are totally
consistent with the absence of the pseudogap region seen in experimental results of optimally as well as
overdoped cuprates.

We next consider the underdoped region with a special emphasis on the resonance peak behavior. 
\begin{figure}[htb]
\epsfxsize=7.8cm
$$\epsffile{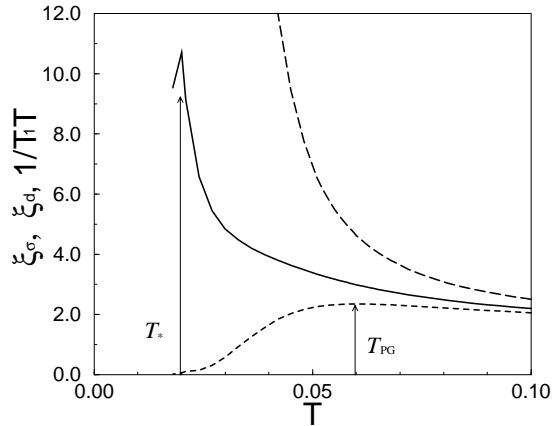}$$
\vspace{-0.7cm}
\caption{Temperature dependence of the spin correlation length (solid line), $1/T_1T$ (dashed line), and the $d$SC correlation length (long-dashed line) in the $\varphi=1$ case. We have taken $|\sGamma|=|\dGamma|=0.7$, $\uss=3.14$, $\udd=3.05$, $\usd=1.0$, $\mu=0.02$ and $t'=-0.02$, which correspond to an underdoped cuprate, YBa${}_2$Cu${}_3$O${}_{6.63}$. As in Fig.~\ref{Figxi-phi0}, the data of $1/T_1T$ are normalized by its value at $T=t$. The fact that $\xid$ starts increasing rapidly with decrease in temperature at $T_*$ can be confirmed by the vanishing amplitude of $1/T_1T\propto\xis^2/(\xis+\xid)$ reached near $T_*$.}
\label{Figxi-phi1}
\end{figure}
The calculated $\xis$, $\xid$ and $1/T_1T$ are shown
in Fig.~\ref{Figxi-phi1} for the parameter values corresponding
to an underdoped cuprates, YBa${}_2$Cu${}_3$O${}_{6.63}$.
Here we take $\varphi=1$. In contrast to the optimally doped case, the spin
correlation length has its maximum at a temperature well above  $T_*$.
 With decrease in temperature below $T_{PG}$, $\xid$ starts
growing quicker than $\xis$.  This competition between $\xid$ and $\xis$ is an origin of the pseudogap formation. Below $T_*$, the $d$SC fluctuations
go into the renormalized-classical regime, which signals the decrease in
$\xis$.  
We again interpret $T_*$ as the rough estimate of $\Tc$.
These properties are also similar to experimental data in underdoped
cuprates with a pseudo spin gap, such as
YBa${}_2$Cu${}_4$O${}_8$~\cite{Y124NMR} and
Bi${}_2$Sr${}_2$CaCu${}_2$O${}_8$~\cite{Bi2212NMR}. 

\begin{figure}[htb]
\epsfxsize=7.8cm
$$\epsffile{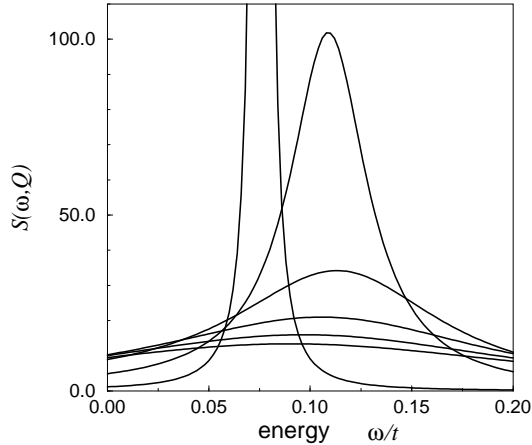}$$
\vspace{-0.7cm}
\caption{The spin structure factor $S(\omega,{\mib Q})$ with ${\mib
Q}=(\pi,\pi)$ for the underdoped case with $\varphi=1$.
From the overdamped side, $T/t=0.069$, $0.06$, $0.051$, $0.042$,
$0.033$ and $0.024$.
The condition $\varphi=1$ makes the spin excitation underdamped
and brings about the shift of the low-energy spectral weights to
higher energies. It evolves as a resonance peak as
observed in neutron scattering experimants.}
\label{FIG_SQw}
\end{figure}

The growth of the pairing correlation  also 
drives reduction of the damping $\gamma_{\sigma}$ in spin excitations and makes  underdamped resonance peak at a finite frequency $\omega=\omega^*$
in $S(Q,\omega)$.  Namely, the spectral weight start transferring from
$\omega=0$ to the peak region around $\omega^{*}$.  A similar crossover was previously obtained
in a numerical calculation near the quantum transition point between
$d$SC and AFM ordered phases~\cite{Assaad96,Assaad97,Assaad97b}.  
The peak structure in $S(Q,\omega)$ around $\omega^*$ reproduces some
qualitative feature in the resonance peak observed
experimentally~\cite{Fong,Y1236.69neutron} as we see in Fig.~\ref{FIG_SQw}.  In our treatment, $\omega^*$
is self-consistently determined from the competition between $d$SC and
AFM and the value $\omega^*$ is characterized by  the $d$SC gap
amplitude.  The AFM fluctuations are pushed out from the region lower
than $\omega^*$ due to the triplet excitation gap generated by the $d$SC gap formation. 

In our framework of (\ref{chi_sigma}),
$\omega^\ast$ has to be proportional to $\xis^{-1}$ for small
$\sgamma$, while experimentally, growth of the  correlation length
takes place with a fixed finite $\omega^\ast$ when the temperature is lowered.  
To reproduce the temperature dependence in experiments, we need to modify the assumed form
(\ref{chi_sigma}) as discussed before~\cite{Assaad97,Onoda991} and  will need to consider  dominat incoherent part in addition to the 
coherent response.

We next discuss the single-particle spectral weight~\cite{Onoda992}. 
Here we compare our results with ARPES data in
Bi2212 with similar values for $\TPG(\sim 170{\rm K})$
and $\Tc(\sim 83{\rm K})$ to those in YBa${}_2$Cu${}_3$O${}_{6.63}$.
We calculate the electronic spectra ${\rm Im} G(\omega,{\mib k})$ using
the same parameter values for the above choice of the underdoped cuprates.
The single-particle Green's function is defined by $G(\omega,{\mib k})=
1/(\omega-\varepsilon({\mib k})-\Sigma(\omega,{\mib k}))$.
Here we calculate the self-energy within the 1-loop level using
\begin{eqnarray}
\lefteqn{{\rm Im}\Sigma(\omega,{\mib k})=
\smint\frac{{\rm d}^2{\mib k}'}{(2\pi)^2}
\int\!\frac{{\rm d}\omega'}{2\pi}{\rm Im} G^{(0)}(\omega',{\mib k}')}
\nonumber\\
&&\times\left[\sGamma^2{\rm Im}\schi(\omega-\omega',{\mib k}-{\mib k}')
\left(\coth\frac{\omega-\omega'}{2T}+\tanh\frac{\omega'}{2T}\right)\right.
\nonumber\\
&&{}+\dGamma^2\frac{g({\mib k})^2+g({\mib k}')^2}{2}
{\rm Im}\dchi(\omega+\omega',{\mib k}+{\mib k}')
\nonumber\\
&&\times\left.\left(\coth\frac{\omega+\omega'}{2T}-\tanh\frac{\omega'}{2T}\right)\right],
\label{Sigma}
\end{eqnarray}
with the bare Green's function $G^{(0)}(\omega,{\mib k})$
and $g({\mib k})=(\cos k_x-\cos k_y)/2$.
Here $\zsdxi{}^{-2}$ has been replaced with $\xi_{\sigma,{\rm d}}^{-2}$.
For the prefactor $A_{\rm d}$ in (\ref{chi_d}),
we take $A_{\rm d}=4t^{-1}$ to give a proper value for the
midpoint shift in ARPES intensity in the pseudogap region~\cite{photoemission}.

\begin{figure}[tb]
\epsfxsize=7.6cm
$$\epsffile{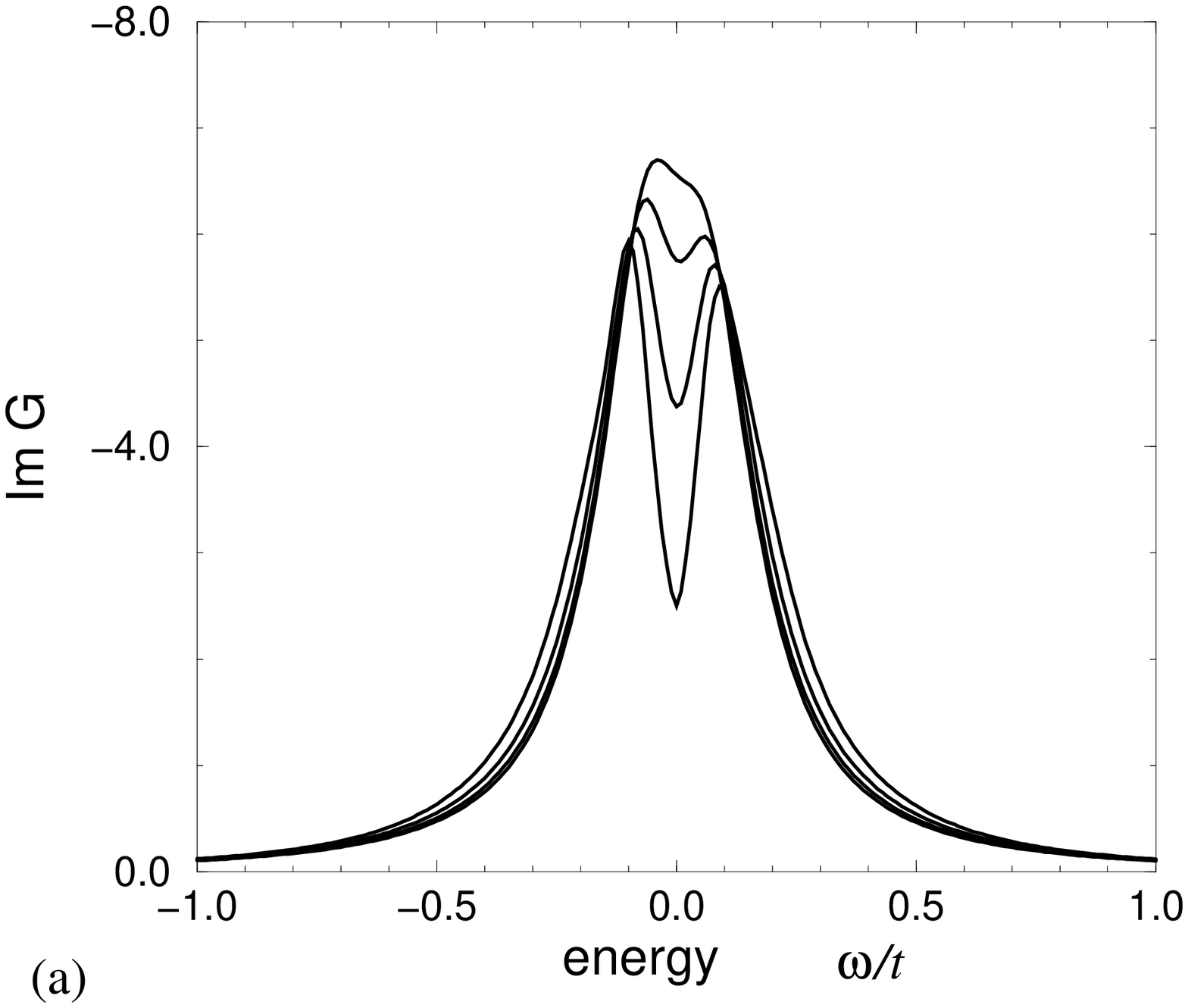}$$
\epsfxsize=7.6cm
$$\epsffile{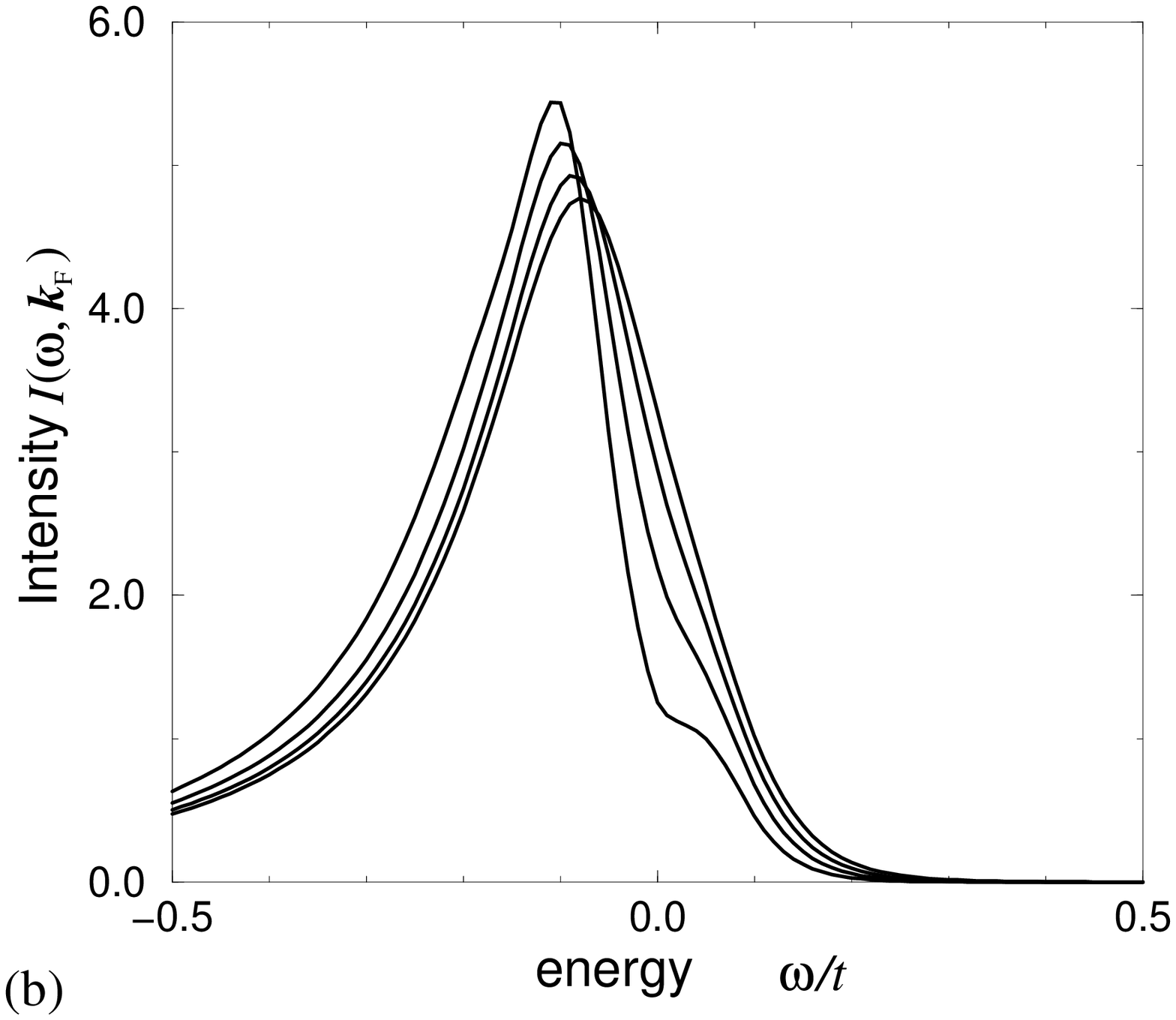}$$
\vspace*{-5mm}
\caption{Temperature dependence of (a) the imaginary part of
  the Green's function and (b) the ARPES intensity for
  ${\mib k}_{\rm F}=(\pi,3\pi/64)$ for $\varphi=1$. This momentum
  point is on the Fermi surface and the closest to $(\pi,0)$ in our
  calculation.
  Temperatures in the plotted data are  $0.069$,
  $0.06$, $0.051$, and $0.042$ in the energy unit of $t$
  from the data with larger intensity at $\omega=0$ both for (a) and 
  (b).}
\label{FIG_p16403}
\end{figure}
Figure~\ref{FIG_p16403}(a) shows ${\rm Im} G(\omega,{\mib k}_{\rm F})$
with ${\mib k}_{\rm F}$ near the flat spot at various temperatures.
At the highest temperature $T\ge 0.069t$, we have a peak at $\omega=0$,
though it is damped by thermal fluctuations.
At lower temperatures still above $\TPG$, only the low-energy spectral
weights gradually start decreasing.
We note that $d$SC correlations grow more
rapidly than those of AFM below $\TPG$ ($\sim 0.06t$)~\cite{Onoda991}. They suppress only
the low-energy part of the peak in the spectral weights.
Well below $\TPG$, ${\rm Im} G$ shows further loss of weights
around $\omega=0$.
For the same momentum ${\mib k}_{\rm F}$,
we also plot the intensities
$I(\omega,{\mib k}_{\rm F})={\rm Im} G(\omega,{\mib k}_{\rm F}) f(\omega)$
in Fig.~\ref{FIG_p16403}(b), where $f$ is
the Fermi function.
The energy of the midpoint is nearly zero at $T=0.06t(\sim\TPG)$.
For $T<\TPG$, the midpoint shifts to higher binding energies.
This shift amounts to $0.045t\sim 11$meV at $T=0.042t(\sim122{\rm K})$,
in agreement with experiments~\cite{photoemission}.

The momentum dependence shows that the low-energy part of the single-particle excitations
is under a stronger suppression near the flat spot ,
while those closer to the nodes are better understood as quasiparticles. 
The calculated results clearly show the formation of the pseudogap first from the $(\pi,0)$ region. The overall result qualitatively well captures the emergence of the pseudogap structure observed in angle-resolved photoemission experiments.

\section{Summary and Discussion} \label{SECTION_summary}

By assuming the d-wave attractive channel and the presence of strongly renormalized flat quasi-particle dispersion around the $(\pi,0)$ region,  we have considered the mode-mode coupling theory for the AFM and $d$SC fluctuations. 
The pseudogap in the high-$\Tc$ cuprates is reproduced as the
region with enhanced $d$SC correlations and is consistently explained from
precursor effects for the superconductivity.  The existence of the flat shoal region plays a role to suppress the effective Fermi temperature $E_F$.  This suppressed $E_F$ and relatively large pairing interaction $\Gamma_d$ both drive the system to the strong coupling region thereby leads to the pseudogap formation.   The pseudogap formation is also enhanced by the AFM fluctuations repulsively coupled with dSC fluctuations.

The pseudogap formation clarified from the interplay of AFM and $d$SC
is summarized as follows:
When the $d$SC correlation grows faster but competes severely with the low-energy AFM fluctuations, the pseudogap structure appears above $T_c$ with a suppression of $T_c$.  The pseudogap is observed clearly in the suppression of $1/T_1T$.  Detailed structure of the pseudogap depends on the damping exponent $\varphi$.  By taking a proper choice of parameters for several underdoped cuprates with $\varphi=1$,
$1/T_1T$ shows a faster decrease at $\TPG(>T_*)$ while $\xis$ continues to increase until $T_*$.  With this parameter values, the resonance peak at a finite frequency in $S(q,\omega)$ is also obtained.  The single-particle spectral weight shows the growth of the gap structure around $(\pi,0)$ and $(0,\pi)$ below $T_{PG}$.
The qualitative similarity between our results for
the underdoped case with $\varphi=1$ and the experimental results in YBa$_2$Cu$_3$O$_{6.63}$ and underdoped Bi$_2$Sr$_2$CaCu$_2$O$_{8+\delta}$ suggests that the damping of
the AFM and $d$SC collective modes decreases in the pseudogap regime at least for these compounds.
It means that low-energy fermions around the flat spots mainly
contribute to the damping.  This is consistent with the strong damping
of quasiparticle around the flat spot observed experimentally in the
underdoped region. 

If the AFM fluctuations are not strong enough and do not compete severely with the dSC fluctuations,  the spin correlation length, $\xis$ and
$1/{}^{63}T_1T$ both reaches its maximum value only at $T=T_*$ and
then decreases as the temperature decreases, which indicates the absence of the pseudogap region.  The experimental results in optimally and overdoped cuprates are reproduced from this choice of parameters together with $\varphi=0$,
namely the case where the damping $\gamma$ does not depend on the $d$SC correlation
length.

The success in reproducing the pseudogap behavior in spin excitations is based on the competition between low-energy AFM and $d$SC fluctuations.
Such competition requires the repulsion for $\usd>0$.
Then the $d$-wave attraction cannot be mediated by low-energy
spin fluctuations.
Although it does not necessarily exclude the attraction generated from the high-energy part of the spin fluctuations,
it requires a formalism for such incoherent contributions beyond the
conventional weak-coupling approach.

Further studies are required for a more complete understanding of
the pseudogap in the high-$\Tc$ cuprates.  Microscopic derivation of our two starting points is the most intriguing future subject.  We
have concentrated on the single-particle excitations only around the
flat spots, $(\pi,0)$ and $(0,\pi)$.  However, in the one-loop
level, the origin of the flat dispersion and strong damping in this
momentum region is not fully clarified.  Experimentally
the flatness and damping strength appear much more pronounced than the
expectation from the one-loop analyses. Numerical analyses also
support that this remarkable momentum dependence around the flat spots
is generated by the  strong correlation effects.  We have to calculate
self-energy corrections as well as the vertex corrections in a
self-consistent fashion to clarify the profoundness of such correlation
effects.  This is clearly the step beyond the one-loop level.   This will also contribute to
clarify how the pairing channel appears and how the flat spots are
destabilized to the paired singlet.  We also note that the dominance
of the incoherent weight over the quasiparticle weight in the
single-particle excitations near the metal-insulator transition may
require a serious modification in the derivation of the AFM and $d$SC
susceptibilities.  The Curie-Weiss type form for the dynamic spin
susceptibility we assumed needs to be reconsidered~\cite{RMP2}, because the spin
susceptibility is also determined mainly from the incoherent part of
the single-particle excitations which we have not considered at all in
this paper.

\acknowledgement
This paper is dedicated to Professor Hiroshi Yasuoka in the occasion of his 60th birthday with special thanks to his leadership for a long time in experimental studies of transition metal oxides.  The work was supported by "Research for the Future" Program from
the Japan Society for the Promotion of Science under the grant number
JSPS-RFTF97P01103.

\newpage


\begin{thebibliography}{99}
\bibitem{RMP}For a recent review see M. Imada, A. Fujimori and Y. Tokura: Rev. Mod. Phys. {\bf 70} (1998) 1039, Sec. IV.C.
\bibitem{LoeserShenDessauMarshallParkFournierKapitulnik1996}
Z.-X. Shen and D. S. Dessau: Physics Reports {\bf 253} (1995) 1;
A. G. Loeser, Z.-X. Shen, D. S. Dessau, D. S. Marshall, C. H. Park, P. Fournier and A. Kapitulnik:
Science {\bf 273} (1996) 325.
\bibitem{photoemission}
H. Ding, T. Yokoya, J. C. Campuzano, T. Takahashi, M. Randeria, M. R. Norman, T. Mochiku, K. Kadowaki and J. Giapintzakis:
Nature {\bf 382} (1996) 51;
\bibitem{Gofron}K. Gofron, J. C. Campuzano, A. A. Abrikosov,
M. Lindroos, A. Bansil, H. Ding, D. Koelling and B. Dabrowski:
Phys. Rev. Lett. {\bf 73} (1994) 3302.
D. S. Marshall, D. S. Dessau, A. G. Loeser, C-H. Park, A. Y. Matsuura,
J. N. Eckstein, I. Bozovic, P. Fournier, A. Kapitulnik,
W. E. Spicer and Z.-X. Shen: Phys. Rev. Lett. {\bf 76} (1996) 4841.
\bibitem{Imada98}M. Imada, F. F. Assaad, H. Tsunetsugu and Y. Motome: cond-mat/9808044 and to be published.
\bibitem{RMPsecIIF11}For a recent review see M. Imada, A. Fujimori and
Y. Tokura: Rev. Mod. Phys. {\bf 70} (1998) 1039, Sec. II.F.11.
\bibitem{Yasuoka}H. Yasuoka, T. Imai and T. Shimizu: ``Strong
Correlation and Superconductivity" ed. by H. Fukuyama, S. Maekawa and
A. P. Malozemoff (Springer Verlag, Berlin, 1989), p.254.
\bibitem{Y124NMR}H. Zimmermann, M. Mali, D.Brinkmann, J. Karpinski, E. Kaldis and S. Rusiecki:
Physica C {\bf 159} (1989) 681;
T. Machi, I. Tomeno, T. Miyataka, N. Koshizuka, S. Tanaka, T. Imai and H. Yasuoka:
Physica C {\bf 173} (1991) 32.
\bibitem{Bi2212NMR}K. Ishida, Y. Kitaoka, K. Asayama, K. Kadowaki and T. Mochiku:
Physica C {\bf 263} (1996) 371.
\bibitem{Hggap}Y. Itoh, T. Machi, A. Fukuoka, K. Tanabe, and
H. Yasuoka: J. Phys. Soc. Jpn. {\bf 65} (1996) 3751.  
\bibitem{Hg1223NMR}M. -H. Julien, P. Carretta, M. Horvati\'{c}, C. Berthier, Y. Berthier, P. S\'{e}gransan, A. Carrington and D. Colson:
Phys. Rev. Lett. {\bf 76} (1996) 4238.
\bibitem{Fong}H. F. Fong, B. Keimer, D. L. Milius and I. A. Aksay:
Phys. Rev. Lett. {\bf 78} (1997) 713.
\bibitem{Assaad98}F. F. Assaad and M. Imada: Eur. Phys. J. B 10 (1999) 595.
\bibitem{Onoda991}S. Onoda and M. Imada: J. Phys. Soc. Jpn. {\bf 68} (1999) 2762.
\bibitem{Onoda992}S. Onoda and M. Imada: J. Phys. Soc. Jpn. {\bf 69} (2000) 312.
\bibitem{RMP2} See ref.~\cite{RMP} Sec.II.D,E,F and IV.C.
\bibitem{SCR}
T. Moriya: {\it Spin Fluctuation in Itenerant Electron Magnetism} (Springer-Verlag, Berlin, 1985).
\bibitem{MerminWagner66}M. D. Mermin and H. Wagner:
Phys. Rev. Lett. {\bf17} (1966) 1133.
\bibitem{K-Ttransition}L. Berezinski: Sov. Phys. JETP {\bf 32} (1970) 493;
J. M. Kosterlitz and D. J. Thouless: J. Phys. C {\bf 6} (1973) 1181.
\bibitem{Hg1201NMR}Y. Itoh, T. Machi, A. Fukuoka, K. Tanabe and
H. Yasuoka: J. Phys. Soc. Jpn. {\bf 65} (1996) 3751.
\bibitem{LSCONMR}T. Imai, K. Yoshimura, T. Uemura, H. Yasuoka and K. Kosuge:
J. Phys. Soc. Jpn. {\bf 59} (1990) 3846.
\bibitem{LSCOARPES}A. Ino, C. Kim, T. Mizokawa, Z.-X. Shen, A. Fujimori, M. Takabe, K. Tamasaku, H. Eisaki and S. Uchida: J. Phys. Soc. Jpn. 68 (1999) 1496;
A. Ino, T. Mizokawa, K. Kobayashi, A. Fujimori, T. Sasagawa, T. Kimura, K. Kishio, K. Tamasaku, H. Eisaki, and S. Uchida: Phys. Rev. Lett. {\bf 81} (1998) 2124;
A. Ino, T. Mizokawa, and A. Fujimori, K. Tamasaku, H. Eisaki, S. Uchida, T. Kimura, T. Sasagawa, and K. Kishio: Phys. Rev. Lett. {\bf 79} (1997) 2101.
\bibitem{Assaad96}F. F. Assaad, M. Imada, and D. J. Scalapino:
Phys. Rev. Lett. {\bf 77} (1996) 4592.
\bibitem{Assaad97}F. F. Assaad, M. Imada, and D. J. Scalapino:
Phys. Rev. B {\bf 56} (1997) 15001.
\bibitem{Assaad97b}F. F. Assaad, and M. Imada:
Phys. Rev. B {\bf 58} (1998) 1845.
\bibitem{Y1236.69neutron}J. Rossat-Mignod, L. P. Renault, C. Vettier, P. Bourges, P. Burlet, J. Bossy, J. Y. Henry and G. Lapertot:
Physica C {\bf 185-189} (1991) 86.
\end{thebibliography}
\end{document}